\documentclass[journal=apchd5,manuscript=article]{achemso}
\usepackage[version=3]{mhchem} % Formula subscripts using \ce{}

\SectionNumbersOn
\usepackage{braket}

\author{Marco Romanelli}
\affiliation{Department of Chemical Sciences, University of Padova, via Marzolo 1, 35131 Padova, Italy}
\altaffiliation{These authors equally contributed to the work}
\author{Rosario R. Riso}
\affiliation{Department of Chemistry, Norwegian University of Science and Technology, 7491 Trondheim, Norway}
\altaffiliation{These authors equally contributed to the work}
\author{Tor S. Haugland}
\affiliation{Department of Chemistry, Norwegian University of Science and Technology, 7491 Trondheim, Norway}
\author{Enrico Ronca}
\affiliation{Department of Chemistry, Biology and Biotechnology, University of Perugia, Via Elce di Sotto, 8, 06123, Perugia, Italy}
\author{Stefano Corni}
\email{ stefano.corni@unipd.it}
\affiliation{Department of Chemical Sciences, University of Padova, via Marzolo 1, 35131 Padova, Italy}
\alsoaffiliation{CNR Institute of Nanoscience, via Campi 213/A, 41125 Modena, Italy}
\alsoaffiliation{Padua Quantum Technologies Research Center, University of Padova, 35131 Padova, Italy}

\author{Henrik Koch}
\email{henrik.koch@sns.it}
\affiliation{Department of Chemistry, Norwegian University of Science and Technology, 7491 Trondheim, Norway}
\alsoaffiliation{Scuola Normale Superiore, Piazza dei Cavalieri 7, 56126 Pisa, Italy}

\title{\textbf{Effective single mode methodology for strongly coupled multimode molecular-plasmon nanosystems}}


\begin{document}

\begin{abstract}
Strong coupling between molecules and quantized fields has emerged as an effective methodology to engineer molecular properties. New hybrid states are formed when molecules interact with quantized fields. Since the properties of these states can be modulated by fine-tuning the field features, 
an exciting and new side of chemistry can be explored. In particular, significant modifications of the molecular properties can be achieved in plasmonic nanocavities, where the field quantization volume is reduced to sub-nanometric volumes. Intriguing applications of nanoplasmonics include the possibility of coupling the plasmons with a single molecule, instrumental for sensing, high-resolution spectroscopy, and single-molecule imaging. In this work, we focus on phenomena where the simultaneous effects of multiple plasmonic modes are critical.
We propose a theoretical methodology to account for many plasmonic modes simultaneously while retaining computational feasibility. Our approach is conceptually simple and allows us to accurately account for the multimode effects and rationalize the nature of the interaction between multiple plasmonic excitations and molecules.   
\end{abstract}

\section{Introduction}
Strong light-matter coupling between molecules and electromagnetic fields lead to the formation of new hybrid states, known as polaritons, where the quantum nature of the electromagnetic field entangles with purely molecular states \cite{fregoniP,bloch2022strongly,flick2017atoms,galego2015cavity,scholes2020polaritons,ramezani2017plasmon,saez2022plexcitonic,sanchez2022theoretical}. The resulting polaritons can display different key features compared to the original states, potentially leading to new chemical/photochemical reactivity \cite{fregoniP,feistP,fregoni2018manipulating,hutchison2012modifying,lather2019cavity,fregoni2020,kowalewski2016,antoniou2020}, energy transfer processes \cite{coles2014,du2018,georgiou2018,zhong2017,saez2018,saez2019} or relaxation channels \cite{munkhbat2018,felicetti2020,antoniou2020,gu2020} among others. While photonic cavities are an obvious choice, other fields, like the ones produced by electronic excitations in plasmonic nanostructure, can also be used to achieve the strong coupling regime. Despite their highly lossy nature, plasmonic nanocavities can confine the electromagnetic fields even down to sub-nanometric volumes \cite{chikkaraddy2016single}. The resulting interaction could be instrumental for a wide range of applications, such as sensing \cite{petryayeva2011,catch2020,etch2008,VODINH1995}, high-resolution spectroscopy \cite{maccaferri2021,haran2010single,zhang2013}, single-molecule imaging \cite{yang2020,zhang2013, jcp2021} and photo-catalysis \cite{cortes2020,zhang2013plasmonic,kale2014direct,zhou2018quantifying,lou2014synthesis}. 
\\
Recent works point out that the simultaneous contribution of multiple plasmonic modes, going beyond the simplest dipolar resonances, might be critical for a number of phenomena\cite{delga2014,feist2021macro,medina2021,sanchez2022few}, e.g. the chiro-optical response of light-matter systems \cite{acsnano2022,MunRho2019,Mun2018,Nanoscale2017,Govorov2012,stevenson2020active,lininger2022chirality,khaliq2021giant}. In such cases, theoretical models that capture multiple plasmons modes simultaneously are of utmost importance. 
\noindent Several \textit{ab initio} quantum electrodynamics (QED) methods for strongly coupled systems have been proposed, e.g. quantum electrodynamics density functional theory (QEDFT) \cite{PhysRevA2014,PhysRevLett2013,flick2017atoms}, QED coupled cluster (QED-CC) \cite{PhysRevX2020, Haugland2021,deprince2021cavity,pavosevic2021polaritonic,riso2022characteristic} and quantum electrodynamics full configuration interaction (QED-FCI)\cite{Haugland2021,riso2022molecular}. Despite its computational affordability, QEDFT inherits the intrinsic problems of exchange and correlation functionals \cite{Flick2018,PhysRevLett2015}, whereas QED-CC, albeit more accurate, is computationally demanding. The latter method has recently been extended to model quantized plasmonic modes obtained through a polarizable continuum model (PCM) \cite{mennucci2012polarizable} description of the nanoparticle response (Q-PCM-NP) \cite{NanoLett2021}.  In its current implementation, however, QED-CC cannot take into account more than one plasmon mode at a time. Generalization of the original theory to the multimode case will quickly become computationally unfeasible.
\noindent In this paper, we couple the existing plasmon QED-CC method to a scheme that captures the main effects of multiple plasmons into a single effective mode. This allows us to retain the same computational cost of a single mode QED-CC calculation while accounting for the multimode effects. \\

This paper is organized as follows. In Section \ref{Theory} the definition of the effective mode is presented. In Section \ref{compdetails} we present the computational details of the numerical examples presented in Section \ref{Results}. Specifically, the effective mode approach is tested on a system composed of 3 nanoparticles (NPs) surrounding either a hydrogen or para-nitroaniline (PNA) molecule. For hydrogen, we benchmark the effective mode approach against multimode QED-FCI. Our final considerations and perspectives are given in Section \ref{Conclusions}. 

\section{Theory}\label{Theory}
\subsection{Effective mode scheme }\label{sub:effmode}
The nanoparticle is described using the Drude-Lorentz dielectric function model for a metallic nanoparticle \cite{jackson1999classical}, that is
\begin{equation}
\epsilon(\omega) = 1 + \frac{\Omega_{P}^{2}}{\omega_{0}^2-\omega^2-i\gamma\omega},
\label{eq:Drude}
\end{equation}
where $\Omega_{P}$ is the bulk plasma frequency, $\omega_{0}$ is the natural frequency of the bound oscillator and $\gamma$ is the damping rate. Together, these quantities define the nanoparticle material. The technique to quantize the NP linear response through a PCM-based theory has already been reported in a previous work \cite{NanoLett2021}.
%\revER{(Is it something missing here??)}
In summary, the nanoparticle surface is described as a discretized collection of tesserae, labeled by $j$, each of which can host a variable surface charge representing the NP response to a given external perturbation\cite{cances1997new,corni2015,Mennucci2019}. The key quantity obtained from the PCM-based quantization scheme is $q_{pj}$ 
which can be identified as the transition charge sitting on the j'th tessera of the NP for a given excited state p. The collection of all the charges for a given p-mode represents one possible normal mode of the NP (a plasmon), with frequency $\omega_p$.
The detailed theory formulation can be found in the original work \cite{NanoLett2021} where the above-mentioned quantities are explicitly derived.
\\
On this basis, the Hamiltonian used to describe the interaction between the nanoparticle and the molecule equals

\begin{equation}
{H} = {H_{e}}+\sum_{p}\omega_{p}b^{\dagger}_{p}b_{p}+\sum_{pj}q_{pj}V_{j}(b^{\dagger}_{p}+b_{p}),  \label{eq:Hamiltonian}
\end{equation}

\noindent where ${H}_{e}$ is the standard electronic Hamiltonian \cite{helgaker2014molecular}, $\omega_{p}$ is the frequency of the p'th nanoparticle mode and the operators ${b}^{\dagger}_{p}$ and ${b}_{p}$ create and annihilate plasmonic excitations of frequency $\omega_{p}$, respectively. The interaction between the molecule and the plasmon is mediated through the bilinear term

\begin{equation}
V = \sum_{pj}q_{pj}V_{j}(b^{\dagger}_{p}+b_{p}).   
\label{eq:biliniear}
\end{equation}

\noindent In Eq.\ref{eq:biliniear}, ${V}_{j}$ is the molecular electrostatic potential operator evaluated at the j'th tessera of the NP while $q_{pj}$ is the quantized charge of mode p that lies on the j'th tessera. From Eq.\ref{eq:biliniear}, the plasmon-molecule coupling for a transition going from the molecular state $S_{0}$ to $S_{n}$ and exciting the plasmon mode p reads
 
\begin{equation}
g_{pn}=\bra{S_n,1_{p}}\sum_{j}q_{pj}{V}_{j}({b}^{\dagger}_{p}+{b}_{p})\ket{S_0,0}=\sum_{j}q_{pj}{V}_{j}^{S_0 \rightarrow S_n},
\label{eq:g}
\end{equation}
where ${V}_{j}^{S_0 \rightarrow S_n}$ is the potential coming from the $S_0 \rightarrow S_n $ transition density at the j'th tessera of the NP surface. The coupling terms in Eq. \ref{eq:g} are the key quantities for simpler approaches to the strong-coupling regime, such as the Jaynes-Cummings (JC) model \cite{JC1963}. This is also the starting point of the effective mode derivation presented in this work. Using the full Hamiltonian in Eq.~\ref{eq:Hamiltonian} is computationally expensive because of the elevated number of plasmon modes that need to be considered. For this reason, it is customary to only include one mode in the Hamiltonian. While the single mode approximation has been used with great success in the past, there are instances where a multimode approach is necessary. One example, for instance, is when multiple plasmonic excitations are almost resonant with the same molecular excitation or, as already discussed in the introduction, when circular dichroism phenomena are studied. To reduce the computational cost while retaining a reasonable accuracy, it would be desirable to define a single effective boson that accounts, on average, for the effect of many modes.   \\ %  
In this framework, the generalization of the single-mode JC Hamiltonian to a multimode plasmonic system is

\begin{equation}
{H}_{JC}^{multi}=\omega_n{\sigma}^{\dagger}{\sigma}+\sum_{p}\omega_{p}{b}^{\dagger}_{p}{b}_{p}+\sum_{p}g_{pn}({b}^{\dagger}_{p}{\sigma}+{b}_{p}{\sigma}^{\dagger}),
\label{eq:multiJC}
\end{equation}
with $\omega_n$ being the frequency of the $S_0 \rightarrow S_n$ excitation and ${\sigma}^\dagger,{\sigma}$ being the molecular raising and lowering operators.
\begin{equation}
\sigma = \ket{S_{0}}\bra{S_{n}} \hspace{1cm} \sigma^{\dagger}= \ket{S_{n}}\bra{S_{0}}.   
\end{equation}
In our case, $\omega_n$ and $g_{pn}$ are the excitation energies and the plasmon-mediated transition coupling elements computed using coupled cluster singles and doubles (CCSD) (more details can be found in Section \ref{compdetails}).
Diagonalization of the Hamiltonian in Eq.~\ref{eq:multiJC} yields the mixed plasmonic-molecular wave functions with corresponding energies. We will simply use the term "polaritonic" to generally refer to those hybrid states from now on even though a mixed plasmon-electronic excitation state is properly called plexciton \cite{NanoLett2021}. In the single mode case, the two eigenstates, typically called Lower and Upper Polaritons (LP,UP), are given by

\begin{equation}
\begin{split}
&\ket{\psi^{LP}}= \ket{S_{n},0}C^{LP}_{mol}+{b}^{\dagger}_1\ket{S_{0},0}C^{LP}_1    \\
&\ket{\psi^{UP}}= \ket{S_{n},0}C^{LP}_1-{b}^{\dagger}_1\ket{S_{0},0}C^{LP}_{mol},
\end{split}   
\label{eq:psiJCone}
\end{equation}
because of the orthogonality constraints.  
\\
On the other hand, the eigenfunctions of the Hamiltonian in Eq.~\ref{eq:multiJC} for the multimode case read

\begin{equation}
\begin{split}
&\ket{\psi^{LP}_{multi}}=\ket{S_{n},0}C^{LP}_{mol}+\sum_{p}{b}^{\dagger}_p\ket{S_{0},0}C^{LP}_p    \\
&\ket{\psi^{UP}_{multi}}=\ket{S_{n},0}C^{UP}_{mol}+\sum_{p}{b}^{\dagger}_p\ket{S_{0},0}C^{UP}_p,   
\end{split}   
\label{eq:psiJCmulti}
\end{equation}
where the coefficients defining the two polaritonic wave functions do not have to satisfy the strict relation in Eq.~\ref{eq:psiJCone}. Moreover, they can be re-written as

\begin{equation}
\begin{split}
\ket{\psi^{LP}_{multi}}&=\ket{S_{n},0}C^{LP}_{mol}+\sum_{p}{b}^{\dagger}_p\ket{S_{0},0}C^{LP}_p \\
& =\ket{S_{n},0}C^{LP}_{mol}+\left(\sum_{p}{b}^{\dagger}_p\ket{S_{0},0}\frac{C^{LP}_p}{\sqrt{\sum_{m}|C^{LP}_m|^2}}\right)\sqrt{\sum_{m}|C^{LP}_m|^2} \\
& =\ket{S_{n},0}C^{LP}_{mol}+\Tilde{b}^{\dagger}_{LP}\ket{S_{0},0}\sqrt{\sum_{m}|C^{LP}_m|^2}\\
\ket{\psi^{UP}_{multi}}&=\ket{S_{n},0}C^{UP}_{mol}+\sum_{p}{b}^{\dagger}_p\ket{S_{0},0}C^{UP}_p \\
& =\ket{S_{n},0}C^{UP}_{mol}+\left(\sum_{p}{b}^{\dagger}_p\ket{S_{0},0}\frac{C^{UP}_p}{\sqrt{\sum_{m}|C^{UP}_m|^2}}\right)\sqrt{\sum_{m}|C^{UP}_m|^2} \\
& =\ket{S_{n},0}C^{UP}_{mol}+\Tilde{b}^{\dagger}_{UP}\ket{S_{0},0}\sqrt{\sum_{m}|C^{UP}_m|^2},
\end{split}
\label{eq:psiLPtrs}
\end{equation}
where the m index labels the plasmon modes. The effective bosons are given by
\begin{equation}
\begin{split}
\Tilde{b}^{\dagger}_{LP}=&\sum_{p}\frac{C^{LP}_p{b}^{\dagger}_p}{\sqrt{\sum_{m}|C^{LP}_m|^2}}   \\
\Tilde{b}^{\dagger}_{UP}=&\sum_{p}\frac{C^{UP}_p{b}^{\dagger}_p}{\sqrt{\sum_{m}|C^{UP}_m|^2}},
\end{split}
\end{equation} 
and have been introduced to describe the photon part of the lower and upper polaritons. The normalization term $\sqrt{\sum_{m}\left|C^{LP/UP}_{m}\right|^{2}}$, is needed to ensure that the effective bosons still respect the commutation relations:
\begin{equation}
\begin{split}
\left[\tilde{b}_{LP},\tilde{b}^{\dagger}_{LP}\right]=1 \;\;and\;\;\left[\tilde{b}_{UP},\tilde{b}^{\dagger}_{UP}\right]=1.
\end{split}
\end{equation}
We point out that, unlike the single mode case in Eq.\ref{eq:psiJCone}, the effective boson operator for the lower and upper polaritons are different from each other. Moreover, the two bosons have a non-zero overlap such that 
\begin{equation}
\left[\tilde{b}_{LP},\tilde{b}^{\dagger}_{UP}\right]\neq 0.    
\end{equation} 
The effective mode approximation comes into play when we seek a single effective mode $\tilde{b}$ that replaces both $\tilde{b}_{LP}$ and $\tilde{b}_{UP}$ such that the energies obtained using the effective upper and lower polaritonic states 

\begin{equation}
\begin{split}
& \ket{\tilde{\psi}^{LP}}=\ket{S_{n},0}C_{mol}+\Tilde{b}^{\dagger}\ket{S_{0},0} \sqrt{\sum_{m}|C_m|^2}  \\
& \ket{\tilde{\psi}^{UP}}=\ket{S_{n},0}\sqrt{\sum_{m}|C_m|^2}-\Tilde{b}^{\dagger}\ket{S_{0},0} C_{mol}
\end{split}   
\label{eq:psiJCeff}
\end{equation}
are as close as possible to the ones obtained using the multimode JC model. We notice that in Eq.\ref{eq:psiJCeff} the $C_{m}$ coefficients are now common to both the lower and upper polariton wavefunctions.
Specifically, they are optimized by minimizing the functional 

\begin{equation}
\begin{split}
f(\mathbf{C})=&\sqrt{\left(\Delta E^{LP}\right)^{2} +\left(\Delta E^{UP}\right)^{2} } \\
\Delta E^{LP/UP}=&\frac{\bra{\tilde{\psi}^{LP/UP}}{H}_{JC}^{multi}\ket{\tilde{\psi}^{LP/UP}}}{\braket{\tilde{\psi}^{LP/UP}|\tilde{\psi}^{LP/UP}}}-E^{LP/UP}_{multi}
\end{split}
\label{eq:functional}
\end{equation}
where $E^{LP}_{multi}$ and $E^{UP}_{multi}$ are the LP and UP energies obtained by diagonalizing the Hamiltonian in Eq.~\ref{eq:multiJC}. We note that the two solutions shown in Eq.~\ref{eq:psiJCeff} resemble the structure of the exact single mode case in Eq.~\ref{eq:psiJCone}. Nonetheless, the plasmonic part of the wave function captures the effect of multiple modes at the same time. The procedure described in this section can easily be generalized to the case of an optical cavity.

\subsection{Transformation to the effective mode basis}\label{sub:unitary}
Once the effective mode has been defined, the Hamiltonian in Eq. \ref{eq:Hamiltonian} can be rewritten as
\begin{equation}
    H = H_{e} + \sum_{pqr} U^{\dagger}_{rp} \omega_{p} U_{pq} \tilde{b}^{\dagger}_{r}\tilde{b}_{q}  + \sum_{pq j}q_{jp}V_{j} U_{pq} (\tilde{b}_{q} + \tilde{b}^{\dagger}_{q}),\label{eq:effective_field_Hamiltonian}
\end{equation}
where $\tilde{b}_{1} = \tilde{b}$ is the effective mode defined in the previous section and the other $\tilde{b}_{q}$ fulfill
\begin{equation}
\left[\tilde{b}_{p},\tilde{b}^{\dagger}_{q}\right]=\delta_{pq}.  \label{eq:Boson}  
\end{equation}
The two bosonic basis are  related by a unitary transformation $U$:
\begin{equation}
\tilde{b}_{p}=\sum_{q}b_{q}U_{qp} \hspace{2cm} U_{q1}=\frac{C_{q}}{\sqrt{\sum_{m}C^{2}_{m}}}.  \label{eq:effective}  
\end{equation}
Truncating the plasmon modes in Eq.~\ref{eq:effective_field_Hamiltonian} to only include the effective mode $\tilde{b}_{1}$, the Hamiltonian reads
\begin{equation}
     H = H_{e} + \tilde{\omega}\tilde{b}^{\dagger}\tilde{b}  + \sum_{j}\tilde{q}_{j}V_{j} (\tilde{b}+\tilde{b}^{\dagger}),\label{eq:Used_Hamiltonian}
\end{equation}
where the subscript $1$ on the plasmon mode has been dropped and the following quantities have been introduced
\begin{equation}
\tilde{\omega}=\sum_{p}U^{\dagger}_{1p}\omega_{p}U_{p1} \hspace{1cm}   \tilde{q}_{j}=\sum_{p}q_{jp}U_{p1}. 
\end{equation} 
The quantized charge $\tilde{q}_{j}$, of the effective plasmon mode $\tilde{b}$, allows for a direct visualization of the effective mode properties (see Fig.\ref{fig:effmode}c).  \\
Starting from the Hamiltonian in Eq.~\ref{eq:Used_Hamiltonian} we can use any single-mode QED method to study the effects of multiple plasmonic modes on molecular properties. In this work, we focus on the QED-CC approach.   
\subsection{QED-CC}\label{sub:QEDCC}
The QED-CC approach is the natural extension of standard coupled cluster theory to the strong coupling regime. The wave function is parametrized as
\begin{equation}
\ket{\psi}= \textrm{exp}\left(T\right)\ket{\rm HF}\otimes\ket{0},    
\end{equation}
where $\ket{\rm HF}$ is the reference Slater determinant (usually obtained through an Hartree-Fock like procedure) while $\ket{0}$ denotes the plasmonic vacuum. The cluster operator $T$ is defined as 
\begin{equation}
\begin{split}
T =& \sum_{ai}t^{a}_{i}E_{ai}+\frac{1}{2}\sum_{aibj}t^{ab}_{ij}E_{ai}E_{bj}+\Gamma b^{\dagger}\\
+&\sum_{ai}s^{a}_{i}E_{ai}b^{\dagger}+\frac{1}{2}\sum_{aibj}s^{ab}_{ij}E_{ai}E_{bj}b^{\dagger},    \label{eq:QED_Cluster}
\end{split}
\end{equation}
with each term corresponding to an electron, electron-plasmon or plasmon excitation. In Eq.~\ref{eq:QED_Cluster}, the electronic second quantization formalism has been adopted such that\cite{helgaker2014molecular}:
\begin{equation}
E_{pq} = \sum_{\sigma}a^{\dagger}_{p\sigma}a_{q\sigma},    
\end{equation}
where $a^{\dagger}_{p\sigma}$ and $a_{q\sigma}$ create and annihilate an electron with spin $\sigma$ in orbitals p and q, respectively. Following the commonly used notation, we denote the unoccupied HF orbitals with the letters $a,b,c...$ while for the occupied orbitals we use $i,j,k...$\cite{helgaker2014molecular}. Inclusion of the full set of excitations in Eq.~\ref{eq:QED_Cluster} leads to the same results as QED-FCI. In this work we truncate $T$ to include up to one plasmon excitation as well as single and double electronic excitations in line with what has been presented in Ref.\cite{PhysRevX2020}. The parameters $t^{a}_{i}, t^{ab}_{ij}, s^{a}_{i}, s^{ab}_{ij}$ and $\Gamma$ are called amplitudes. They are determined solving the projection equations
\begin{equation}
\Omega_{\mu,n} = \bra{\mu,n} e^{-T}\bar{H}e^{T}\ket{\rm{HF},0},   
\label{eq:Projection}
\end{equation}
where $\mu$ is an electronic excitation while $n$ is a plasmonic excitation. We adopted the notation
\begin{equation}
\ket{\mu,n}=\ket{\mu}\otimes\ket{n}. \label{eq:projection}   
\end{equation}
The $\bar{H}$ operator is the molecule-plasmon Hamiltonian in Eq.~\ref{eq:Used_Hamiltonian} transformed with a coherent state. This accounts for the polarization of the plasmonic system induced by the molecular charge density in the HF state.
\begin{equation}
\bar{H}=e^{-z(\tilde{b}-\tilde{b}^{\dagger})}He^{z(\tilde{b}-\tilde{b}^{\dagger})} \hspace{2cm} z = -\frac{1}{\tilde{\omega}}\bra{\rm HF}\sum_{j}\tilde{q}_{j}V_{j}\ket{\rm HF}.     
\end{equation}\begin{figure}[ht!]
    \centering
    \includegraphics[width=1.0\textwidth]{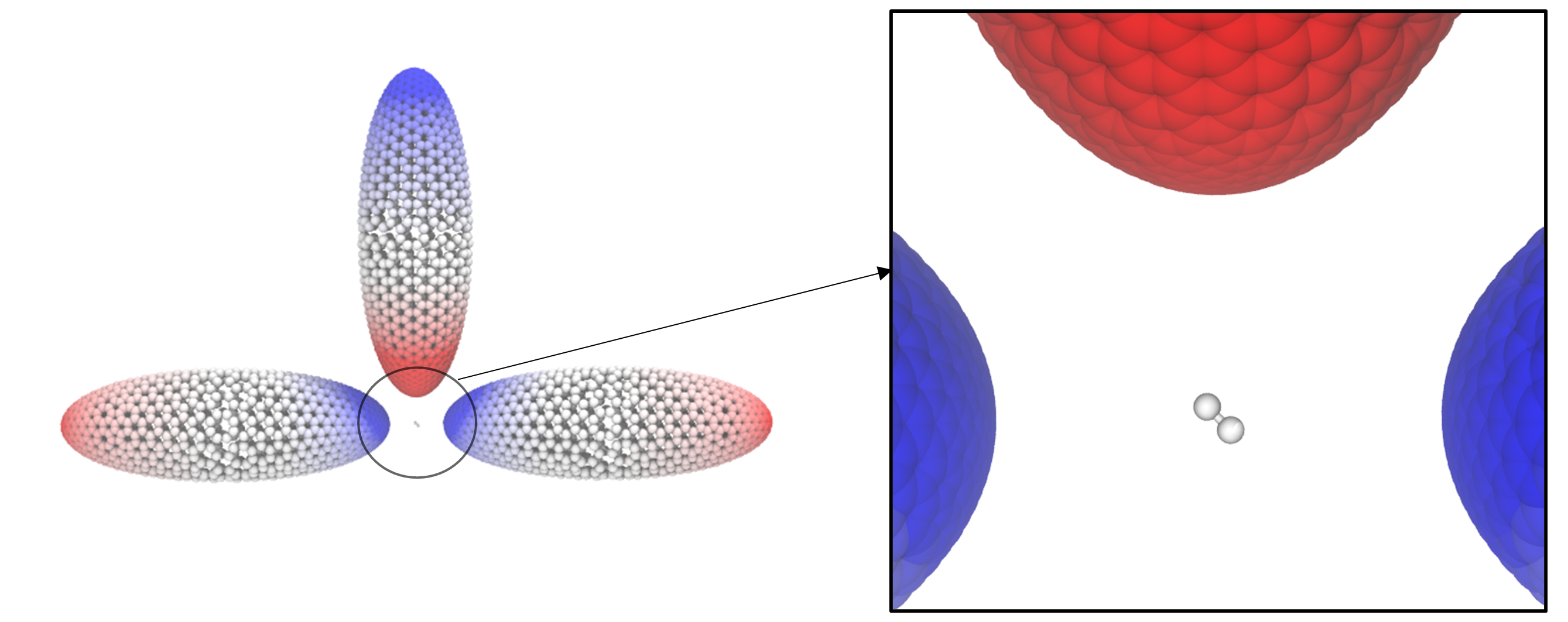}
    \caption{ Setup employed to test the effective mode scheme. The plasmonic system consists of 3 ellipsoidal NPs surrounding an $H_2$ molecule in the yz plane. The beads composing the NPs represent the centroids of each tessera upon surface discretization and host a given quantized charge $q_{pj}$. The lowest (in energy) plasmon mode is shown, red beads refer to positive charges, whereas blue ones refer to negative charges.  Each NP is $\approx$ 0.6 nm far from $H_2$.}
    \label{fig:setup}
\end{figure}
\begin{figure}[ht!]
    \centering
    \includegraphics[width=1.0\textwidth]{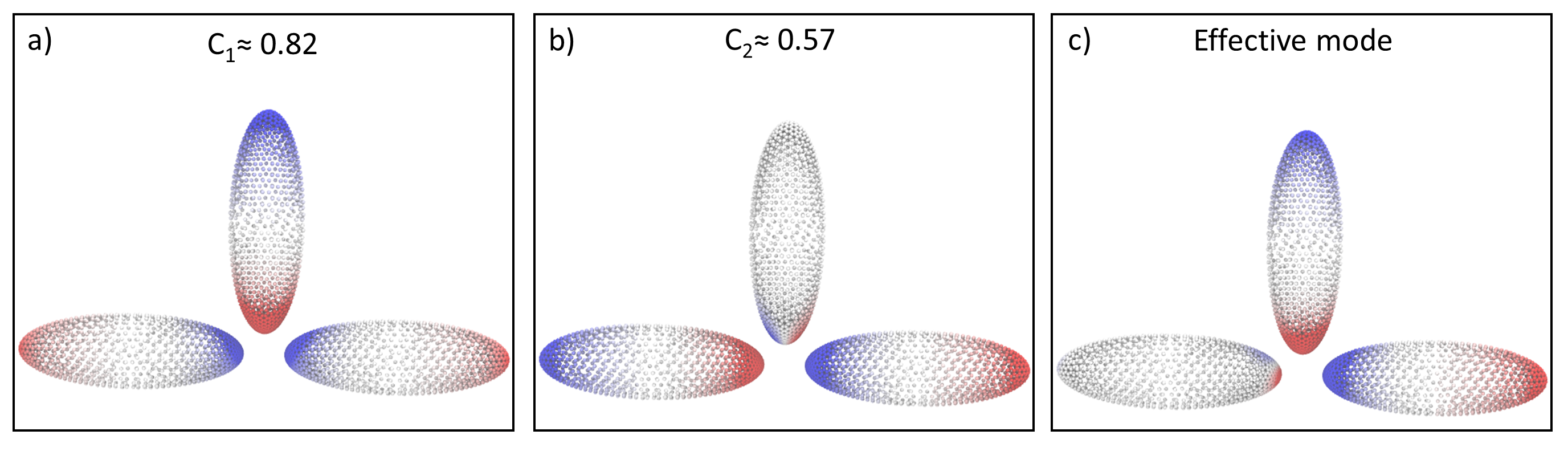}
    \caption{ First (a), and second (b), quasi-degenerate plasmon modes of the setup shown in Fig. \ref{fig:setup}. The energy splitting between these two modes is $\approx$ 16 meV and they both significantly couple to the $S_0 \rightarrow S_1$ transition of $H_2$. Their contribution to the effective mode are shown at the top of the panel. c) Visualization of the optimized effective mode. Only the two most important modes are reported in panels a) and b), but the first 12 modes coupling to the molecular transition, contribute to the effective mode optimization. }
    \label{fig:effmode}
\end{figure}

\section{Computational details}\label{compdetails}
The setup we employed to test the effective mode approach consists of three identical ellipsoidal NPs, each one featuring a long-axis length of 6.0 nm and a short-axis length of 2.0 nm. In between the nanoellipses we placed an $H_2$ molecule that is approximately 0.6 nm away from the three structures, as shown in Fig.\ref{fig:setup}. The surface meshes were created using the Gmsh code\cite{gmsh} and consist of $\approx$ 4500 tesserae. The Drude-Lorentz parameters used to define the metal dielectric function are $\Omega_{P}=8.605$ eV, $\gamma = 0.217$ eV and $\omega_{0}=12.517$ eV, which is very close to the value adopted for silver in previous literature works \cite{Zeman1987}. The damping rate and the natural frequency of the oscillator were chosen such that the NPs lowest plasmon mode is resonant with the $H_2$ $S_0 \rightarrow S_1$ molecular transition, around 
12.7 eV according to CCSD/aug-cc-pVDZ calculations in vacuum \cite{pritchard2019a,dunning1989a,kendall1992a}. In order to avoid fictitious charge transfer effects between the nanoparticles, an \textit{a posteriori} charge normalization scheme, described in Ref.~\cite{jcp2021}, has been applied to each NP.  
\\ A similar setup was used for testing the effective mode performance with the PNA molecule. This time $\omega_0$ was set to $4.354$ eV such that the first plasmonic modes match the first transition of PNA at approximately 4.8 eV, according to CCSD/cc-pVDZ \cite{pritchard2019a,dunning1989a,kendall1992a}. The electronic calculations were performed using a development version of the eT program \cite{folkestad2020t}.

\begin{figure}[htbp]
    \centering
    \includegraphics[width=1.0\textwidth]{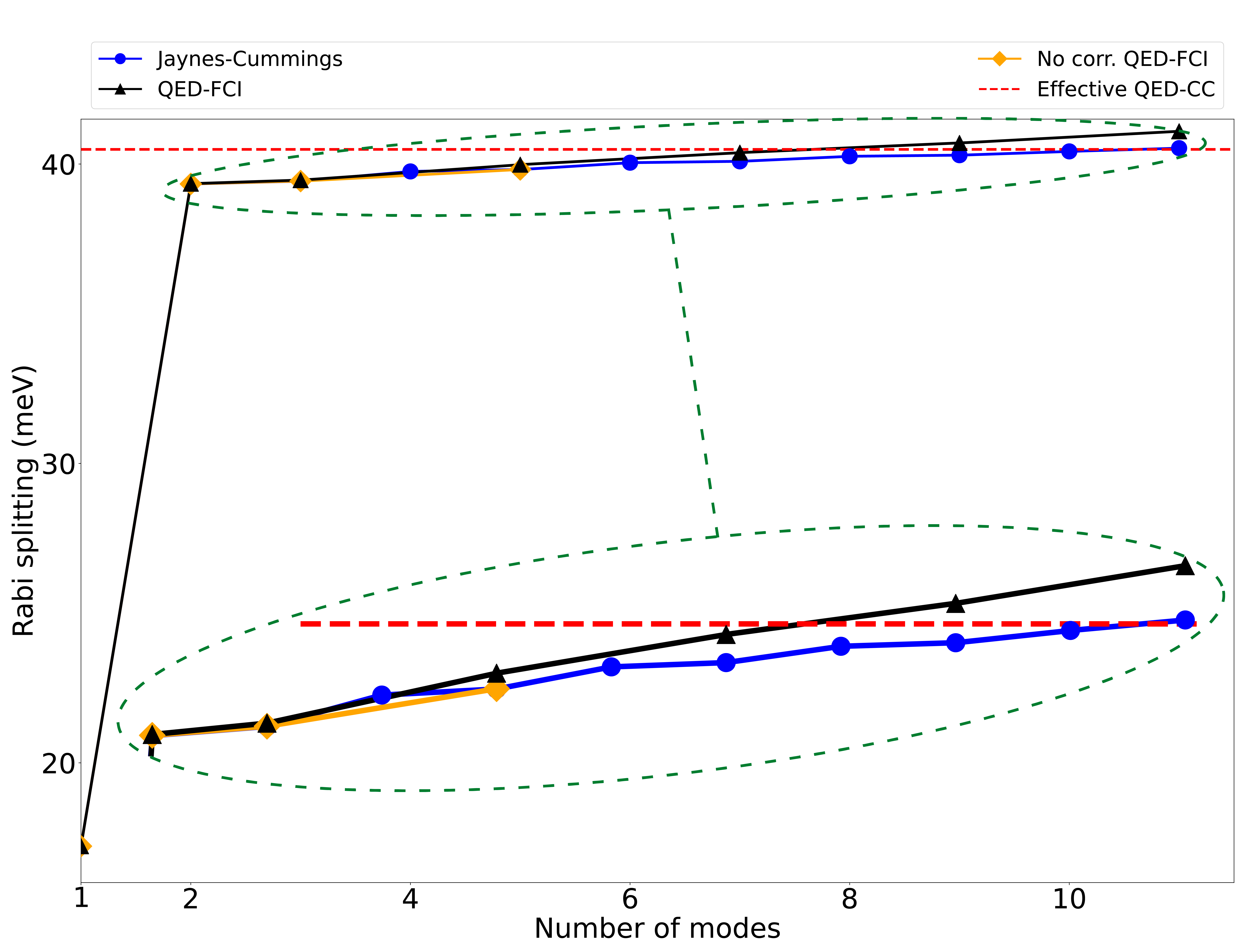}
    \caption{Computed Rabi splitting for the setup shown in Fig.\ref{fig:setup} as a function of the number of plasmonic modes included in the Hamiltonian. The calculations have been perfomed with the following methods: multimode Jaynes-Cummings, QED-FCI and QED-FCI without relaxation of the electronic wave function (No corr. QED-FCI). We highlight that the latter curve is basically overlapped with the blue one. The effective mode QED-CC (dashed red line) recovers most of the multiphoton contribution. We note that the effective mode optimization has been computed using the first 12 plasmon modes of the nanoparticle setup whose coupling with the molecular transition is non-zero.}
    \label{fig:rabi}
\end{figure}

\begin{figure}
    \centering
    \includegraphics[width=\textwidth]{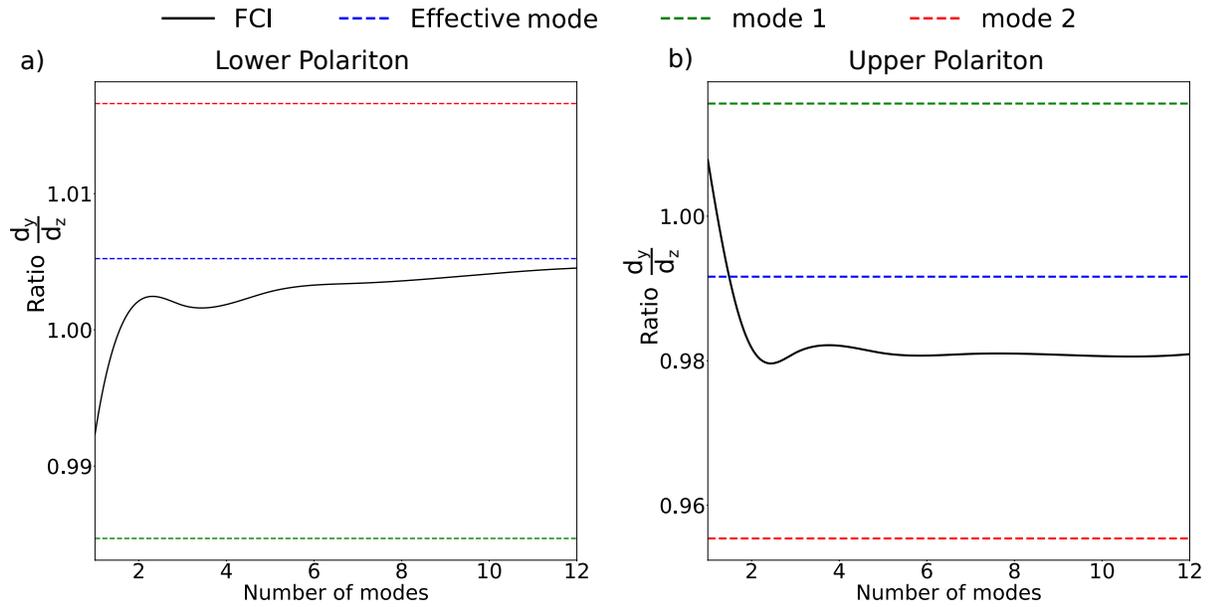}
    \caption{Ratio of the y and z component of the LP(a)/ UP(b) transition dipole for the setup of Fig.\ref{fig:setup} in 4 different cases: QED-CC using mode 1 (dashed green line) or mode 2 only (dashed red line), effective mode QED-CC (dashed blue line) and QED-FCI (solid black curve). The QED-FCI data are reported as a function of the number of plasmonic modes included in the Hamiltonian, up to 12, which corresponds to the maximum number of modes employed in the effective mode optimization.}
    \label{fig:dip_tr}
\end{figure}

\section{Results and discussion}\label{Results}
The NPs setup shown in Fig.\ref{fig:effmode} has two almost degenerate low excitations (Fig.\ref{fig:effmode} a and b) at 12.661 eV and 12.677 eV, whose coupling parameters with the first $H_{2}$ transition are 8.6 meV and 15.2 meV, respectively.
Both excitations will significantly contribute to the effective mode. Specifically, their coefficients in the expansion of the effective mode, see Eq.\ref{eq:effective} are reported in the top part of Fig.\ref{fig:effmode}. \\
In Fig.\ref{fig:rabi} we show how the inclusion of multiple plasmon modes affects the $H_{2}$ Rabi splitting. Results are shown for the multimode JC Hamiltonian, QED-FCI and the effective mode approach for QED-CC. We notice that, as expected, the single mode approximation underestimates the Rabi splitting by almost a factor two. All the multimode methods therefore show a large improvement once the second mode has been added. Inclusion of additional modes still enlarges the splitting, although we note that the change is quite small when compared to the improvement observed adding the second plasmon in the picture. Despite using a single bosonic operator, the effective mode QED-CC allows us to almost exactly capture the multimode effect with a predicted Rabi splitting of 40.49 meV compared to 41.09 meV (QED-FCI value). We notice that the QED-FCI and JC results are not exactly equal and that the error increases when more modes are considered. This difference is due to the relaxation of the electronic ground and excited states induced by the presence of the nanoparticle. This effect is not captured unless an \textit{ab initio} approach is used. If the electronic wave function is not optimized in the QED-FCI calculations, thus not accounting for the mutual polarization with the NP (the no corr. QED-FCI in Fig.\ref{fig:rabi}), the difference between QED-FCI and JC dramatically reduces. Nonetheless, the differences between the \textit{ab initio} method and the two level approximation are small when compared to the improvement from 1 to 2 modes.
\\
We also investigated other excited state properties, like the molecular contribution to the transition dipole moment in the $GS \rightarrow LP/UP$ transition. As shown in Fig.\ref{fig:dip_tr}a-\ref{fig:dip_tr}b, the ratio between the molecular y and z components (the $H_2$ molecule lies in the yz plane) of the polaritonic transition dipole approaches the exact QED-FCI limit when the effective mode is used. On the other hand, the agreement is significantly worse using either mode 1 or mode 2 separately. This shows that the effective mode not only improves the Rabi splitting description compared to the single-mode approximation, but also provides a better description of the most important excited state properties, e.g. transition dipoles/densities.

%PNA

\begin{figure}[ht!]
    \centering
    \includegraphics[width=1.0\textwidth]{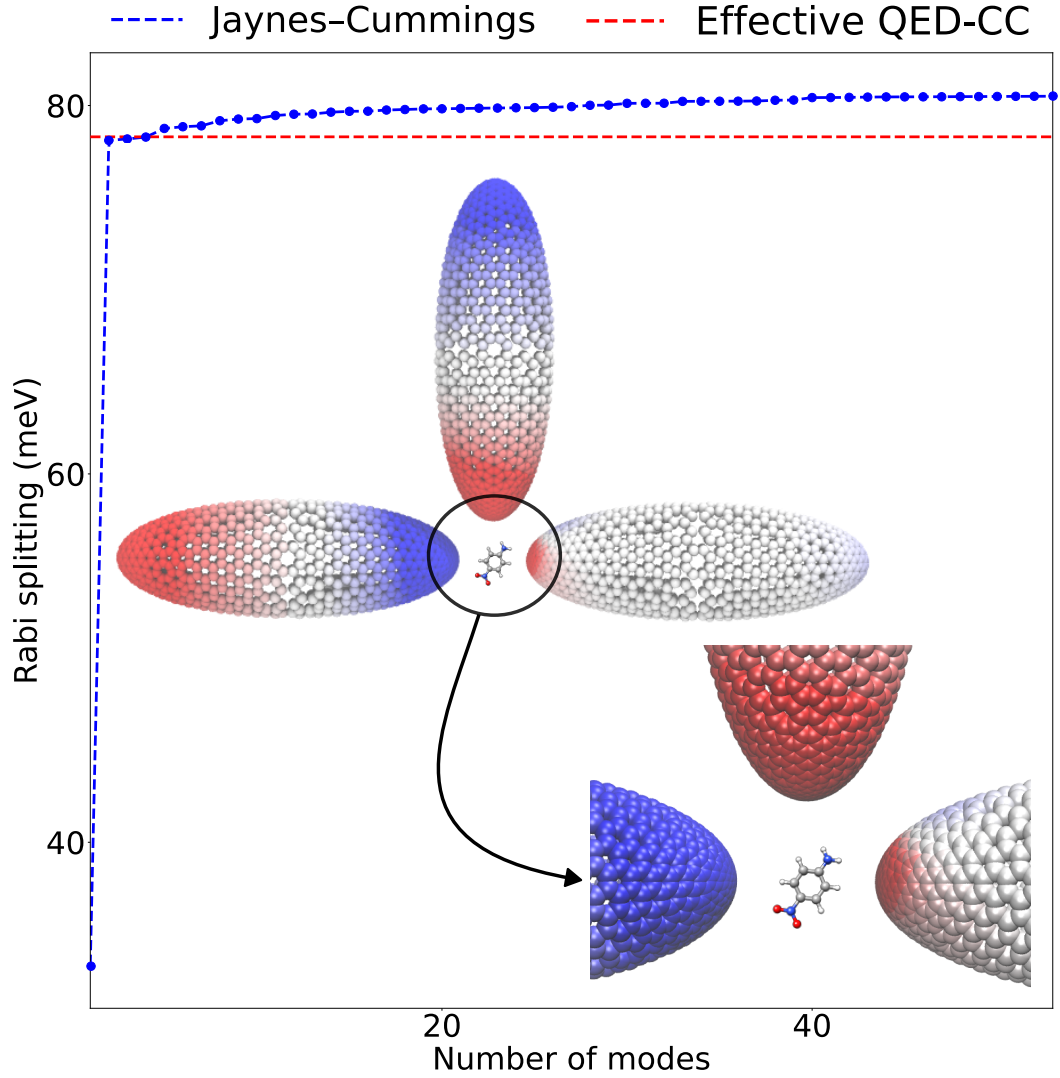}
    \caption{Computed Rabi splitting for a setup of 3 ellipsoidal NPs surrounding a PNA molecule. The average distance from the metallic surfaces is $\approx$ 0.6 nm (setup shown as inset). The dashed blue line are the results obtained through the multimode Jaynes-Cummings Hamiltonian (see Eq.\ref{eq:multiJC}), whereas the red one is the outcome of the QED-CC effective mode approach. Visualization of the optimized effective mode is shown in the inset.}
    \label{fig:PNA}
\end{figure}
\begin{figure}[ht!]
    \centering
    \includegraphics[width=1.0\textwidth]{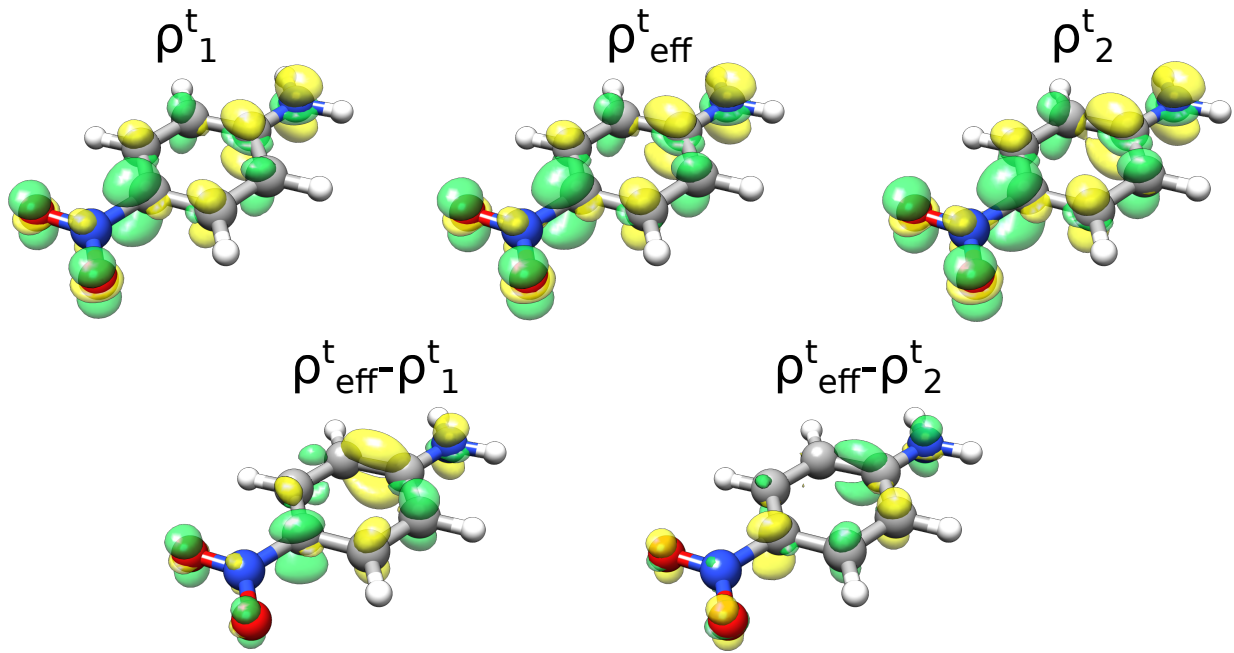}
    \caption{PNA transition density plots for the  $GS\rightarrow LP$ transition (see setup of Fig.\ref{fig:PNA}), computed using QED-CC with mode 1 and 2 or the effective mode. Positive density contributions are reported in yellow, whereas negative ones are reported in green. The difference between the transition density obtained with the effective mode and either mode 1 or mode 2 is also shown, thus allowing an easier visualization of the major changes in the PNA transition density upon changing the plasmonic part of the QED-CC Hamiltonian. }
    \label{fig:tr_dens}
\end{figure}
\noindent The qualitative picture does not change if a more complicated molecule like paranitroaniline (PNA) is placed between the three nanoellipsoidal structures (see Fig.\ref{fig:PNA}). Comparing the multimode JC results with the effective mode approach for PNA, we notice that, similarly to the $H_{2}$ case, the effective mode QED-CC recovers most of the multimode contribution. Specifically, the Rabi splitting predicted by QED-CC is almost the same as using 5 field modes in the JC approach (78.3 meV). In Fig.\ref{fig:tr_dens}, we compare the $GS \rightarrow LP$ transition densities of PNA when either the effective mode or a single mode approach is used. Notably, an enhancement of the LP charge transfer character can be observed moving from the single mode QED-CC with the lowest plasmon mode ( $\rho^{\textrm{t}}_{1}$ of Fig.\ref{fig:effmode}a) to the effective mode QED-CC, $\rho^{\textrm{t}}_{\textrm{eff}}$. The difference between the two transition densities, $\rho^{\textrm{t}}_{\textrm{eff}}-\rho^{\textrm{t}}_{1}$, indeed shows an increased negative density contribution on the $NO_2$ group (acceptor) and an increased positive density contribution on the $NH_2$ group (donor). The opposite trend is observed in the case of mode 2 (Fig.\ref{fig:effmode}b), meaning that in the mode 2 case more charge is transferred compared to the effective mode case. These findings can be easily rationalized using the theory described in Sec.\ref{Theory}. Indeed, since mode 2 is favouring the charge separation more than mode 1, the effective plasmon, that is a linear combinations of mode 1 and mode 2, predicts an intermediate transfer between the two.
Since an increasing number of modes are coupled with the main molecular transition, nanoplasmonic systems with multiple almost degenerate excitations represent a promising option to increase the field effects without reducing the field quantization volume. 

\section{Conclusions}\label{Conclusions}
Building on the previously developed Q-PCM-NP/QED-CC model\cite{NanoLett2021}, we propose here a framework to account for multimode environments using a single effective mode. Our approach captures the main features arising from the simultaneous coupling to multiple plasmons, while retaining the same computational cost of single mode methods \cite{PhysRevX2020,NanoLett2021}. Physical quantities, such as Rabi splittings and transition dipoles, are correctly reproduced, as verified by benchmarking against exact multimode QED-FCI for the hydrogen molecule surrounded by 3 ellipsoidal nanoparticles. The same theoretical approach is applied to a larger organic molecule, para-nitroaniline (PNA), where QED-FCI or multimode calculations are out of reach. Our results demonstrate that the inclusion of multiple modes is critical to correctly evaluate the plasmon-matter interaction in the case of quasi-degenerate plasmonic modes. In these cases, indeed, the single-mode approximation naturally breaks down. We notice that the effective mode scheme can be applied to any kind of wave function approximation and is not specific for plasmonic systems. We also point out that the effective mode is optimized to correctly reproduce the upper and lower polaritons only. Therefore, no improvement in the description of the ground state should be expected with this methodology. A generalization of the method should, however, be able to model the effect of multiple plasmonic modes on the molecular ground state. This topic will be the subject of a future publication. 
\\ As a number of \textit{ab initio} QED methods have started to appear recently \cite{PhysRevA2014,PhysRevLett2013,flick2017atoms,PhysRevX2020, Haugland2021,deprince2021cavity,pavosevic2021polaritonic,riso2022molecular,mandal2020polarized}, the here-developed effective mode approach will be of great use in all those cases where multimode effects need to be taken into account, while retaining a computationally feasible methodology \cite{acsnano2022,MunRho2019,Mun2018,Nanoscale2017,Govorov2012}.  

\section{Acknowledgement}
R.R.R, T.S.H and H.K. acknowledge funding from the Research Council of Norway through FRINATEK Project No. 275506. This work has received funding from the European Research Council (ERC) under the European Union’s Horizon 2020 Research and Innovation Programme (grant agreement No.101020016). M.R acknowledges MIUR “Dipartimenti di Eccellenza” under the project Nanochemistry for energy and Health (NExuS) for funding the Ph.D. grant.
%\bibliography{achemso-demo}
\providecommand{\latin}[1]{#1}
\makeatletter
\providecommand{\doi}
  {\begingroup\let\do\@makeother\dospecials
  \catcode`\{=1 \catcode`\}=2 \doi@aux}
\providecommand{\doi@aux}[1]{\endgroup\texttt{#1}}
\makeatother
\providecommand*\mcitethebibliography{\thebibliography}
\csname @ifundefined\endcsname{endmcitethebibliography}
  {\let\endmcitethebibliography\endthebibliography}{}

\end{document}